\documentclass[10pt, letterpaper]{article}
\usepackage[utf8]{inputenc}
\usepackage{amsmath,amssymb,amsfonts}
\usepackage{graphicx}
\usepackage{mathabx}
\usepackage{verbatim}
\usepackage{geometry}
\usepackage{amsmath,amssymb,amsfonts}
\usepackage{graphicx}
\usepackage{mathabx}
\usepackage{verbatim}
\usepackage{geometry}
\usepackage{bm}

\newcommand\keyword[1]{\textit{\textbf{Keywords---}}\textbf{#1}}

\title{Local Observability of VINS and LINS}
\author{Xinran Li $ ^{1,2}$
\thanks{$^{1}$ Aerospace Information Research Institute, Chinese Academy of Sciences, Beijing 100190, China}%
\thanks{$^{2}$ School of Electronic, Electrical and Communication Engineering, University of Chinese Academy of Sciences, Beijing 100049, China}
\thanks{E-mail addresses:
	{\tt\small lixinran19@mails.ucas.ac.cn}}
}

\date{}
\begin{document}
	\maketitle
	
	\begin{abstract}
		This work analyzes unobservable directions of Vision-aided Inertial Navigation System (VINS) and Lidar-aided Inertial Navigation System (LINS) nonlinear model. Under the assumption that there exist two features observed by the camera without occlusion, the unobservable directions of VINS are uniformly globally translation and global rotations about the gravity vector. The unobservable directions of LINS are same as VINS, while only one feature need to be observed. Also, a constraint in Observability-Constrained VINS (OC-VINS) is proved.
	\end{abstract}
	
	\keyword{Vision-aided inertial navigation, Lidar-aided inertial navigation, observability analysis}
	
	\section{INTRODUCTION}
	
	\ \  \ \  \ \  Observability of VINS and LINS is an important issue concerning about simultaneous localization and mapping (SLAM). Linearlized systems have been investigated by a number of studies.$^{[1], \ [2]}$ The analysis of the unobservable directions of linearlized systems improves the accuracy and avoid degeneracy in Extended Kalman Filter (EKF) and Error-state Kalman Filter (ESKF). 
	
	Hesch and  Roumeliotis have analyzed the nonlinear model of VINS,$^{[3]}$ however, according to Isidori,$^{[4]}$ the nonsingular condition of the codistribution is not satisfied. Theorems of local observability does not hold in the entire state space. However, under the assumption that there exist two features observed by the camera without occlusion, the nonsingular condition is satisfied.
	
	Moreover, the local observability analysis of LINS is given in this paper, using the same method as that of VINS. At last, a proof of Observability-Constrained VINS (OC-VINS) is given.
	
	\section{SYSTEM MODEL}
	
	\ \  \ \  \ \ The state vector is that $\bm{\bar{x}}=\left[\bm{\theta}, \bm{b}_g, \bm{v}, \bm{b}_a, ^G\bm{p}_I | ^G\bm{p}_{f_1}, ... ,^G\bm{p}_{f_N}\right]$, where $\bm{\theta}$ is the Lie algebra of the global rotation of IMU, $\bm{b}_g$ is the bias of the gyroscope, $\bm{v}$ is the global velocity of IMU, $\bm{b}_a$ is the bias of the accelerometer, $^G\bm{p}_I$ is the global position of IMU and  $^G\bm{p}_{f_i}$ is the global position of the i-th feature.

	Using the  Cayley–Gibbs–Rodriguez parameterization, the state vector is 
	\begin{center}
		$\bm{\Psi}(\bm{\bar{x}})=\left[\bm{s}(\bm{\theta}), \bm{b}_g, \bm{v}, \bm{b}_a, ^G\bm{p}_I | ^G\bm{p}_{f_1}, ... ,^G\bm{p}_{f_N}\right]$
	\end{center} where $\bm{\Psi}$ is diffeomorphism. Let $^G\bm{p}_f$ represent any of a feature, the kinematic model is 
	
	\begin{equation}
			\begin{bmatrix}
			\dot{\bm{s}}	\\
			\dot{\bm{b}_g}	\\
			\dot{\bm{v}}	\\
			\dot{\bm{b}_a}	\\
			\dot{^G\bm{p}_I}	\\
			\dot{^G\bm{p}_f}	
		\end{bmatrix}=
		\underbrace{\begin{bmatrix}
				-\frac{\partial \bm{s}}{\partial \bm{\theta}}\bm{b_g}	\\
				\bm{0}	\\
				\bm{g}-\bm{C}^T \bm{b}_a	\\
				\bm{0}	\\
				\bm{v}	\\
				\bm{0}	
		\end{bmatrix}}_{f_0}+
		\underbrace{\begin{bmatrix}
				\frac{\partial \bm{s}}{\partial \bm{\theta}}	\\
				\bm{0}	\\
				\bm{0}	\\
				\bm{0}	\\
				\bm{0}	\\
				\bm{0}	
		\end{bmatrix}}_{f_1} \bm{\omega}+
		\underbrace{ \begin{bmatrix}
				\bm{0}	\\
				\bm{0}	\\
				\bm{C}^T	\\
				\bm{0}	\\
				\bm{0}	\\
				\bm{0}	
		\end{bmatrix}}_{f_2} \bm{a}
	\end{equation} where 
	\begin{equation}
		\frac{\partial \bm{s}}{\partial \bm{\theta}} = \frac{1}{2} (\bm{I}+\bm{ss}^T+\left[\bm{s} \times\right])
	\end{equation}

	In VINS, the observation is 
	\begin{equation}
		\bm{h} = \frac{1}{p_z} \begin{pmatrix}
			p_x\\
			p_y
		\end{pmatrix}
	\end{equation} where $^I\bm{p}_f = \bm{C}(^G\bm{p}_f- \ ^G\bm{p}_I) = \begin{pmatrix}
	p_x \\ p_y \\ p_z
	\end{pmatrix}$ is the position of the feature in the IMU frame.
	
	While in LINS, the observation is 
	\begin{equation}
		\bm{h} = \ ^I\bm{p}_f
	\end{equation}
	\section{OBSERVABILITY OF VINS}
	\ \  \ \  \  The Lie derivatives are
	
	\begin{equation}
		\bm{O} = \begin{bmatrix}
			d\bm{h}  \\ dL_{f_1^i}\bm{h} \\dL_{f_0}\bm{h} \\	dL_{f_1^i}L_{f_1^j}\bm{h}\\ 	dL_{f_0}L_{f_1^j}\bm{h} \\dL_{f_2^i}L_{f_0}\bm{h}\\dL_{f_1^i}L_{f_0}\bm{h} \\ dL_{f_0}L_{f_0}\bm{h} \\ ...
		\end{bmatrix}	
	\end{equation}
	
	$\textbf{Lemma 3.1.}$ Let $null(\bm{A})$ be the right nullspace of $\bm{A}$.
	\begin{equation}
		null(\bm{L} \	\begin{bmatrix}
			d\bm{h}  \\ dL_{f_1^i}\bm{h} \\dL_{f_0}\bm{h} \\	dL_{f_1^i}L_{f_1^j}\bm{h}\\ 	dL_{f_0}L_{f_1^j}\bm{h} \\dL_{f_2^i}L_{f_0}\bm{h}\\dL_{f_1^i}L_{f_0}\bm{h} \\ dL_{f_0}L_{f_0}\bm{h} \\ ... \\ dL_{f_1^k}L_{f_0}L_{f_0} \bm{h}
		\end{bmatrix}	)=null(\begin{bmatrix}
			d\bm{h}  \\ dL_{f_1^i}\bm{h} \\dL_{f_0}\bm{h} \\	dL_{f_1^i}L_{f_1^j}\bm{h}\\ 	dL_{f_0}L_{f_1^j}\bm{h} \\dL_{f_2^i}L_{f_0}\bm{h}\\dL_{f_1^i}L_{f_0}\bm{h} \\ dL_{f_0}L_{f_0}\bm{h} \\ ... \\ dL_{f_1^k}L_{f_0}L_{f_0} \bm{h}
		\end{bmatrix})
	\end{equation}where $\bm{L}$ is a lower triangular matrix, whose diagonal elements are 1.
	
	Using Lemma 1, the Lie derivatives can be simplified by linear combination.

	Now, we calculate the Lie derivatives recursively.
	
	\begin{equation}
		d\bm{h}=\frac{1}{p_z} \begin{bmatrix}
			1& 0 & -\frac{p_x}{p_z} \\
			0 & 1 & -\frac{p_y}{p_z}
		\end{bmatrix}
		\begin{bmatrix}
			\left[^I\bm{p}_f \times \right]\frac{\partial \bm{\theta}}{\partial \bm{s}},\bm{0},\bm{0},\bm{0},-\bm{C},\bm{C}
		\end{bmatrix}
	\end{equation}
	
	Denote $\bm{H_c}=\frac{1}{p_z} \begin{bmatrix}
		1& 0 & -\frac{p_x}{p_z} \\
		0 & 1 & -\frac{p_y}{p_z}
	\end{bmatrix}$, $\bm{K}=\begin{bmatrix}
		\left[^I\bm{p}_f \times \right]\frac{\partial \bm{\theta}}{\partial \bm{s}},\bm{0},\bm{0},\bm{0},-\bm{C},\bm{C}
	\end{bmatrix}$ 
	
		$\textbf{Lemma 3.2.}$ $\forall \bm{v}$,  \begin{equation}
		d\bm{H}_c \bm{v}= -\frac{1}{p_z}(\bm{H}_c \bm{v})  \bm{e}_3^T \bm{K}-\frac{\bm{e}_3^T \bm{v}}{p_z}d\bm{h}
	\end{equation}
	
	proof. $d\bm{H}_c \bm{v} = \begin{bmatrix}\bm{v}^T	\begin{bmatrix}
			0 & 0 &- \frac{1}{p_z^2} \\
			0&0&0\\
			-\frac{1}{p_z^2}& 0& 2 \frac{p_x}{p_z^3}
		\end{bmatrix} 
		\vspace{2pt}
		
		\\
		
		\bm{v}^T	\begin{bmatrix}
			0&0&0 \\
			0&0&-\frac{1}{p_z^2} \\
			0&-\frac{1}{p_z^2}&2\frac{p_y}{p_z^3}
		\end{bmatrix}
	\end{bmatrix} \bm{K} =  \begin{bmatrix}\bm{v}^T	\begin{bmatrix}
			0 & 0 &- \frac{1}{p_z^2} \\
			0&0&0\\
			0& 0&  \frac{p_x}{p_z^3}
		\end{bmatrix} 
		\vspace{2pt}
		
		\\
		
		\bm{v}^T	\begin{bmatrix}
			0&0&0 \\
			0&0&-\frac{1}{p_z^2} \\
			0&0&\frac{p_y}{p_z^3}
		\end{bmatrix}
	\end{bmatrix} \bm{K} +  \begin{bmatrix}\bm{v}^T	\begin{bmatrix}
			0 & 0 &0 \\
			0&0&0\\
			-\frac{1}{p_z^2}& 0&  \frac{p_x}{p_z^3}
		\end{bmatrix} 
		\vspace{2pt}
		
		\\
		
		\bm{v}^T	\begin{bmatrix}
			0&0&0 \\
			0&0&0 \\
			0&-\frac{1}{p_z^2}&\frac{p_y}{p_z^3}
		\end{bmatrix}
	\end{bmatrix} \bm{K}$
	\vspace{2pt}
	
	$=-\frac{1}{p_z}\begin{bmatrix}
		\bm{0}_{2 \times 2}& \bm{H}_c \bm{v}
	\end{bmatrix}\bm{K} -  \frac{\bm{e}_3^T \bm{v}}{p_z}\bm{H}_c\bm{K}$ = $-\frac{1}{p_z}(\bm{H}_c \bm{v})  \bm{e}_3^T \bm{K}-\frac{\bm{e}_3^T \bm{v}}{p_z}d\bm{h}$
	
	$\hfill \Box$
	
	Denote
	\begin{equation}
		\begin{matrix}
			\overline{ dL_{f_1^i} \bm{h} } 	= dL_{f_1^i}\bm{h}+\frac{\bm{e}_3^T \left[^I\bm{p}_f \times \right]\bm{e}_i}{p_z}d\bm{h}		
			
			\\
			
			=-\frac{1}{p_z} (\bm{H}_c \left[^I\bm{p}_f \times \right]\bm{e}_i)  \bm{e}_3^T \bm{K}-\bm{H}_c \left[\bm{e}_i \times \right] \bm{K}	\end{matrix}		\end{equation} and
		
		\begin{equation}
			\begin{matrix}
				\overline{ dL_{f_0} \bm{h} }=  dL_{f_0} \bm{h} + \bm{b}_g^i\overline{ dL_{f_1^i} \bm{h} } -\frac{\bm{e}_3^T (\bm{Cv}+\left[^I\bm{p}_f \times \right]\bm{b}_g)}{p_z}d\bm{h}
				\\
				=dL_{f_0} \bm{h} + \bm{b}_g^i dL_{f_1^i}\bm{h} -\frac{\bm{e}_3^T \bm{Cv}}{p_z}d\bm{h}
			\vspace{2pt}	\\
				=-\bm{H}_c \bm{G} -\bm{H}_c \left[^I\bm{p}_f \times \right] \bm{J} + \frac{1}{p_z}(\bm{H}_c \bm{Cv})  \bm{e}_3^T \bm{K}
		\end{matrix} \end{equation} Define $\bm{G}=\begin{bmatrix}
		\left[\bm{Cv} \times \right]\frac{\partial \bm{\theta}}{\partial \bm{s}},\bm{0},\bm{C},\bm{0},\bm{0},\bm{0}
	\end{bmatrix}$, $\bm{J}=\begin{bmatrix}
	\bm{0},\bm{I},\bm{0},\bm{0},\bm{0},\bm{0}
\end{bmatrix}$. In this paper, we use Einstein summation convention.

		\begin{equation}
			\begin{matrix}
				\overline{dL_{f_2^i}L_{f_0} \bm{h}} = d(\overline{ dL_{f_0} \bm{h} } \ f_2^i) -\frac{\bm{e}_3^T}{p_z}\bm{H}_c \bm{e}_i d\bm{h}\\
				=dL_{f_2^i}L_{f_0} \bm{h}-\frac{\bm{e}_3^T}{p_z}\bm{H}_c \bm{e}_i d\bm{h}\\
				=\frac{1}{p_z} (\bm{H}_c \bm{e}_i)  \bm{e}_3^T \bm{K}
			\end{matrix}
		\end{equation}
		
		$\textbf{Theorem 3.1.}$ Denote $\bm{n} \in null(O)$, then $\bm{Kn} = \bm{0}$
		
		proof. Using Lemma 2, $\overline{dL_{f_2^i}L_{f_0} \bm{h}} \ \bm{n} = \bm{0}$, i.e.
		
		\begin{center}
			$\frac{\bm{e}_3^T \bm{Kn} }{p_z} (\bm{H}_c \bm{e}_i) \ (\forall i)$
		\end{center} $\therefore \bm{e}_3^T \bm{Kn} = 0$. Let $\bm{Kn} = \begin{pmatrix}
		k_1 \\ k_2 \\ 0
		\end{pmatrix}$. $\because d\bm{h\ n}=0$
		
		$\therefore$
		
		\begin{center}
			$\bm{H}_c \begin{pmatrix}
				k_1 \\ k_2 \\ 0
			\end{pmatrix} = \bm{0}$
		\end{center}
		
		$\because p_z \ne 0$, $\therefore k_1 = k_2 = 0$. $\therefore \bm{Kn} = \bm{0}$
		
		$\hfill \Box$
		
		$\textbf{Corollary 3.1.}$ If there are N features, denote $\bm{K}_i=\begin{bmatrix}
			\left[^I\bm{p}_{f_i} \times \right]\frac{\partial \bm{\theta}}{\partial \bm{s}},\bm{0},\bm{0},\bm{0},-\bm{C},\bm{0}_{3 \times 3(i-1)},\bm{C}, \bm{0}_{3 \times 3(N-i)}
		\end{bmatrix}$ and $\bm{n} = \left[\bm{n}_1,\bm{n}_2,\bm{n}_3,\bm{n}_4,\bm{n}_5 \ | \ \bm{n}_{f_1},...\bm{n}_{f_N}\right]^T \in null(\bm{O}) $, $\bm{K}_i \bm{n} =\bm{0} \ (\forall i)$.
		
		Next $\bm{Gn}$ and $\bm{Jn}$ will be calculated.
		
		\begin{equation}
			\begin{matrix}
				\overline{dL_{f_1^k}L_{f_0} \bm{h}} = d(\overline{ dL_{f_0} \bm{h} } \ f_1^k) - (dL_{f_1^i} \bm{h} \ f_1^k)d(\bm{b}_g^i) + (d\bm{h} f_1^k) d(\frac{\bm{e}_3^T \bm{Cv}}{p_z}) \\
				+\frac{\bm{e}_3^T}{p_z}(-\left[\bm{Cv} \times \right] e_k +\frac{\bm{e}_3^T \left[ ^I\bm{p}_f \times \right] e_k}{p_z}\bm{Cv})d\bm{h} \\
				=dL_{f_1^k}L_{f_0} \bm{h} + \bm{b}_g^i dL_{f_1^k}L_{f_1^i} \bm{h} -\frac{\bm{e}_3^T \bm{Cv}}{p_z}dL_{f_1^k} \bm{h}
				+\frac{\bm{e}_3^T}{p_z}(-\left[\bm{Cv} \times \right] e_k +\frac{\bm{e}_3^T \left[ ^I\bm{p}_f \times \right] e_k}{p_z}\bm{Cv})d\bm{h} 
				\vspace{2pt}
				\\
				=(\bm{H}_c \left[\bm{e}_k \times \right] + \frac{\bm{e}_3^T \left[ ^I\bm{p}_f \times \right] e_k}{p_z} \bm{H}_c + \frac{\bm{H}_c \left[ ^I\bm{p}_f \times \right] \bm{e}_k \bm{e}_3^T}{p_z})\bm{G} 
				\vspace{2pt}
				\\
				+(\bm{H}_c\left[\bm{e}_i \times \right] \left[ ^I\bm{p}_f \times \right] \bm{e}_k+\frac{\bm{e}_3^T \left[ ^I\bm{p}_f \times \right] \bm{e}_k}{p_z} \bm{H}_c \left[ ^I\bm{p}_f \times \right] \bm{e}_i+\frac{\bm{e}_3^T \left[ ^I\bm{p}_f \times \right] \bm{e}_i}{p_z} \bm{H}_c \left[ ^I\bm{p}_f \times \right] \bm{e}_k)\bm{e}_i^T\bm{J} \\
				+\bm{T}(\bm{Cv},^I\bm{p}_f)\bm{K}
			\end{matrix}
		\end{equation} where 
		\begin{center}
			$\bm{T}(\bm{Cv},^I\bm{p}_f) = \frac{1}{p_z}\bm{H}_c \left[\bm{Cv} \times \right] \bm{e}_k \bm{e}_3^T - \frac{2\bm{e}_3^T \left[ ^I\bm{p}_f \times \right] \bm{e}_k}{p_z^2}\bm{H}_c \bm{Cv} \bm{e}_3^T -\frac{1}{p_z}\bm{H}_c \bm{Cv} \bm{e}_3^T \left[\bm{e}_k \times \right] -\frac{\bm{e}_3^T \bm{Cv}}{p_z^2}\bm{H}_c \left[ ^I\bm{p}_f \times \right]  \bm{e}_k \bm{e}_3^T$
		\end{center}\ 
		
		$\textbf{Theorem 3.2.}$ When there are two features satisfying that $^I\bm{p}_{f_1} \ne k \ ^I\bm{p}_{f_2}$ ($k\in R$), $\bm{n}\in null(\bm{O})$ satisfies that $\left[\bm{Cv} \times \right]\frac{\partial \bm{\theta}}{\partial \bm{s}} \bm{n}_1 +\bm{Cn}_3 = \bm{0}$ and $\bm{n}_2=\bm{0}$.
		
		proof. $\because \overline{ dL_{f_0} \bm{h} } \ \bm{n} = \bm{0}$ \  $\therefore \left[\bm{Cv} \times \right]\frac{\partial \bm{\theta}}{\partial \bm{s}} \bm{n}_1 +\bm{Cn}_3 + \left[ ^I\bm{p}_f \times \right] \bm{n}_2 = s^I\bm{p}_f \ (s \in R)$.
		
		$\because \overline{dL_{f_1^k}L_{f_0} \bm{h}} \ \bm{n}=\bm{0}$ \  
		
		$\therefore$ 
		\begin{center}
			$\bm{H}_c \left[\bm{e}_k \times \right] \bm{Gn} + \bm{H}_c\left[\bm{Jn} \times \right] \left[ ^I\bm{p}_f \times \right] \bm{e}_k +(\frac{\bm{e}_3^T \left[ ^I\bm{p}_f \times \right] e_k}{p_z} \bm{H}_c + \frac{\bm{H}_c \left[ ^I\bm{p}_f \times \right] \bm{e}_k \bm{e}_3^T}{p_z})(\left[^I\bm{p}_f \times \right]\bm{J} + \bm{G}) \bm{n} = \bm{0} (\forall k)$
		\end{center}

		$\therefore$  \begin{center}
		$-\bm{H}_c \left[\bm{Gn} \times \right] \bm{e}_k + \bm{H}_c\left[\bm{Jn} \times \right] \left[ ^I\bm{p}_f \times \right] \bm{e}_k+s\bm{H}_c \left[ ^I\bm{p}_f \times \right] \bm{e}_k  = \bm{0} (\forall k)$
		\end{center}
		
		$\therefore$ \begin{center}
			$ -\bm{H}_c \left[\bm{Gn} \times \right]  + \bm{H}_c\left[\bm{Jn} \times \right] \left[ ^I\bm{p}_f \times \right] +s\bm{H}_c \left[ ^I\bm{p}_f \times \right]   = \bm{0}$
		\end{center}  
		
		Append $^I\bm{p}_f$, it can be seen that
		
		$-\bm{H}_c \left[\bm{Gn} \times \right]^I\bm{p}_f = \bm{0}$, i.e. 
		\begin{center}
		$\left[ ^I\bm{p}_f \times \right]\left[ ^I\bm{p}_f \times \right] (\left[\bm{Cv} \times \right]\frac{\partial \bm{\theta}}{\partial \bm{s}} \bm{n}_1 +\bm{Cn}_3 ) = 0$
		\end{center} 
		
		$\because rank (\left[ ^I\bm{p}_f \times \right]\left[ ^I\bm{p}_f \times \right]) = 2$ \ $\therefore \left[\bm{Cv} \times \right] \bm{n}_1 +\bm{Cn}_3 = w ^I\bm{p}_f (w \in R)$, i.e. 
		\begin{center}
			$\left[^I\bm{p}_f \times \right](\left[\bm{Cv} \times \right]\frac{\partial \bm{\theta}}{\partial \bm{s}} \bm{n}_1 +\bm{Cn}_3) = \bm{0}$
		\end{center}
		
		$\therefore \left[^I\bm{p}_f \times \right]\bm{n}_2 = (s-w) ^I\bm{p}_f$ \  $\therefore 0=(s-w) (^I\bm{p}_f)^T \ ^I\bm{p}_f$ \  
		
		$\therefore$
		\begin{center}
			$ \left[^I\bm{p}_f \times \right]\bm{n}_2 = \bm{0}$
		\end{center} 
		
		Obviously, $\left[\bm{Cv} \times \right]\frac{\partial \bm{\theta}}{\partial \bm{s}}\bm{n}_1+\bm{C} \bm{n}_3$ and $\bm{n}_2$ is independent on $\bm{n}_{f_k}$ and $^I\bm{p}_{f_k}$. When there are two features satisfying that $^I\bm{p}_{f_1} \ne k \ ^I\bm{p}_{f_2}$, $rank \begin{pmatrix}
			\left[^I\bm{p}_{f_1} \times \right] \vspace{1pt}\\
			\left[^I\bm{p}_{f_2} \times \right]
		\end{pmatrix} =3$, so $\left[\bm{Cv} \times \right]\frac{\partial \bm{\theta}}{\partial \bm{s}} \bm{n}_1 +\bm{Cn}_3 = \bm{0}$ and $\bm{n}_2=\bm{0}$.
		
		$\hfill$ $\Box$
		
		Denote $\bm{N}=\left[\bm{0},\bm{0},\bm{0},\bm{I},\bm{0},\bm{0}\right]$ and $\bm{M}=\left[  \left[\bm{Cg} \times \right]\frac{\partial \bm{\theta}}{\partial \bm{s}},\bm{0},\bm{0},\bm{0},\bm{0},\bm{0} \right]$
		
		\begin{equation}
			\begin{matrix}
				\overline{dL_{f_0}L_{f_0} \bm{h}} = d(\overline{ dL_{f_0} \bm{h} } \ f_0) - (dL_{f_1^i} \bm{h} \ f_0)d(\bm{b}_g^i) + (d\bm{h} f_0) d(\frac{\bm{e}_3^T \bm{Cv}}{p_z}) \\
				+\frac{\bm{e}_3^T}{p_z}(\left[\bm{Cv} \times \right] \bm{b}_g +\bm{b}_a-\bm{Cg}-\frac{\bm{e}_3^T(	\left[^I\bm{p}_{f} \times \right] \bm{b}_g + \bm{Cv})}{p_z}\bm{Cv})d\bm{h} \\
				
				=dL_{f_0}L_{f_0} \bm{h}+\bm{b}_g^i dL_{f_0}L_{f_1^i} \bm{h} -\frac{\bm{e}_3^T \bm{Cv}}{p_z}dL_{f_0} \bm{h} 
				+\frac{\bm{e}_3^T}{p_z}(\left[\bm{Cv} \times \right] \bm{b}_g +\bm{b}_a-\bm{Cg}-\frac{\bm{e}_3^T(	\left[^I\bm{p}_{f} \times \right] \bm{b}_g + \bm{Cv})}{p_z})d\bm{h} \\
				
				=\bm{H}_c(\bm{N}-\bm{M}) \\
				+\bm{T}_1(^I\bm{p}_{f},\bm{b}_g,\bm{Cv})\bm{G}+\bm{T}_2(^I\bm{p}_{f},\bm{b}_g,\bm{Cv})\bm{J} 				+\bm{T}_3(^I\bm{p}_{f},\bm{b}_g,\bm{Cv},\bm{Cg},\bm{b}_a)\bm{K}
		\end{matrix}	\end{equation}where \begin{center}
		$\begin{matrix}
			\bm{T}_1(^I\bm{p}_{f},\bm{b}_g,\bm{Cv})=-\bm{H}_c(\left[\bm{b}_g \times \right]+\frac{2\bm{Cv}\bm{e}_3^T}{p_z}+\frac{ \left[^I\bm{p}_{f} \times \right] \bm{b}_g \bm{e}_3^T }{p_z}+\frac{\bm{e}_3^T(	\left[^I\bm{p}_{f} \times \right] \bm{b}_g + \bm{Cv})}{p_z}\bm{I})  \vspace{4pt}\\
			\bm{T}_2(^I\bm{p}_{f},\bm{b}_g,\bm{Cv}) = -\bm{H}_c(\frac{\bm{Cv}}{p_z}\bm{e}_3^T\left[^I\bm{p}_{f} \times \right]-\left[\bm{Cv} \times \right])
			\vspace{2pt}\\
			-\bm{H}_c(\left[\bm{e}_i \times \right]+\frac{1}{p_z}\left[^I\bm{p}_{f} \times \right]\bm{e}_i\bm{e}_3^T-\frac{\bm{e}_3^T\left[^I\bm{p}_{f} \times \right]\bm{e}_i}{p_z}\bm{I})(\left[^I\bm{p}_{f} \times \right] \bm{b}_g+\bm{Cv})\bm{e}_i^T \vspace{4pt} \\
			\bm{T}_3(^I\bm{p}_{f},\bm{b}_g,\bm{Cv},\bm{Cg},\bm{b}_a) =\bm{H}_c \left[\frac{2\bm{e}_3^T(\left[^I\bm{p}_{f} \times \right] \bm{b}_g+\bm{Cv})}{p_z^2}\bm{Cv} +\frac{\bm{e}_3^T\bm{Cv}}{p_z^2} \bm{I} \right] \bm{e}_3^T 
			\vspace{2pt}
			\\
			-\frac{1}{p_z}\bm{H}_c(\left[\bm{Cv} \times \right] \bm{b}_g +\bm{b}_a-\bm{Cg})\bm{e}_3^T
		\end{matrix}$
	\end{center}
	
		\begin{equation}
		\begin{matrix}
			\overline{dL_{f_1^k}L_{f_0}L_{f_0} \bm{h}} = d(\overline{dL_{f_0}L_{f_0} \bm{h}} \ f_1^k) -(dL_{f_0}L_{f_1^i} \bm{h} \ f_1^k)d\bm{b}_g^i -(dL_{f_0} \bm{h} \ f_1^k)d(\frac{\bm{e}_3^T \bm{Cv}}{p_z}) \\
			-(d\bm{h} \ f_1^k)d(\frac{\bm{e}_3^T}{p_z}(\left[\bm{Cv} \times \right] \bm{b}_g +\bm{b}_a-\bm{Cg}-\frac{\bm{e}_3^T(	\left[^I\bm{p}_{f} \times \right] \bm{b}_g + \bm{Cv})}{p_z})) \\
			=dL_{f_1^k}L_{f_0}L_{f_0} \bm{h}+\bm{b}_g^i dL_{f_1^k}L_{f_0}L_{f_1^i} \bm{h}-\frac{\bm{e}_3^T \bm{Cv}}{p_z}dL_{f_1^k}L_{f_0} \bm{h} \\
			+\frac{\bm{e}_3^T}{p_z}(\left[\bm{Cv} \times \right] \bm{b}_g +\bm{b}_a-\bm{Cg}-\frac{\bm{e}_3^T(	\left[^I\bm{p}_{f} \times \right] \bm{b}_g + \bm{Cv})}{p_z})dL_{f_1^k} \bm{h} \\
			=\bm{H}_c\left[\bm{e}_k \times \right]\bm{M}-\frac{\bm{H}_c}{p_z}( \left[^I\bm{p}_f \times \right]\bm{e}_k\bm{e}_3^T + \bm{e}_3^T\left[^I\bm{p}_f \times \right] \bm{e}_k\bm{I})(\bm{N}-\bm{M}) 	\\
			+\bm{T'}_1\bm{G}+\bm{T'}_2\bm{J} 				+\bm{T'}_3\bm{K}
		\end{matrix}
	\end{equation}
	
	$\textbf{Theorem 3.3.}$ When there are two features satisfying that $^I\bm{p}_{f_1} \ne k \ ^I\bm{p}_{f_2}$ ($k\in R$), $\bm{n}\in null(\bm{O})$ satisfies that $\left[\bm{Cg} \times \right]\frac{\partial \bm{\theta}}{\partial \bm{s}} \bm{n}_1 = \bm{0}$ and $\bm{n}_4=\bm{0}$.
	
	proof. $\because \overline{dL_{f_0}L_{f_0} \bm{h}}  \ \bm{n} = \bm{0}$ \ $\therefore -\left[\bm{Cg} \times \right]\frac{\partial \bm{\theta}}{\partial \bm{s}} \bm{n}_1 +\bm{n}_4 = s \ ^I\bm{p}_f \ (s \in R)$
	
	$\because \overline{dL_{f_1^k}L_{f_0}L_{f_0} \bm{h}}  \ \bm{n} = \bm{0}$
	
	$\therefore$
	\begin{center}
		$ -\bm{H}_c\left[\bm{M}\bm{n} \times \right]\bm{e}_k - s\bm{H}_c \left[^I\bm{p}_f \times \right]\bm{e}_k =\bm{0} \ (\forall k)$
	\end{center}
	
	$\therefore$
	\begin{center}
	$\begin{matrix}
			\bm{H}_c\left[\bm{M}\bm{n} \times \right]+s\bm{H}_c \left[^I\bm{p}_f \times \right]=\bm{0}
		\end{matrix}$
	\end{center}
	Append $^I\bm{p}_f$, $\bm{H}_c\left[^I\bm{p}_f \times \right]\bm{M}\bm{n}=\bm{0}$, i.e.
	
	\begin{center}
		$\left[ ^I\bm{p}_f \times \right]\left[ ^I\bm{p}_f \times \right]\left[\bm{Cg} \times \right]\frac{\partial \bm{\theta}}{\partial \bm{s}} \bm{n}_1 = \bm{0}$
	\end{center}

	$\therefore$
	
	\begin{center}
		$\begin{matrix}
			\left[^I\bm{p}_f \times \right]\left[\bm{Cg} \times \right] \frac{\partial \bm{\theta}}{\partial \bm{s}}\bm{n}_1 = \bm{0} \vspace{2pt}
			\\
			\left[^I\bm{p}_f \times \right]\bm{n}_4 = \bm{0}
		\end{matrix}$
	\end{center}
	
	Because $\left[\bm{Cg} \times \right]\frac{\partial \bm{\theta}}{\partial \bm{s}}\bm{n}_1$ and $ \bm{n}_4$ are independent on $\bm{n}_{f_k}$ and $^I\bm{p}_{f_k}$, when there are two features satisfying that $^I\bm{p}_{f_1} \ne k \ ^I\bm{p}_{f_2}$, $\left[\bm{Cg} \times \right] \frac{\partial \bm{\theta}}{\partial \bm{s}}\bm{n}_1 = \bm{0}$ and $\bm{n}_4=\bm{0}$.
	
	$\hfill$ $\Box$
	
	$\textbf{Theorem 3.4.}$ When there exist two features satisfying that $^I\bm{p}_{f_1} \ne k \ ^I\bm{p}_{f_2}$, $\bm{O \ n} = \bm{0} \iff$
	\begin{center}
		$\begin{bmatrix}
			\left[^I\bm{p}_{f_1} \times \right]\frac{\partial \bm{\theta}}{\partial \bm{s}} & \bm{0} & \bm{0}& \bm{0} &-\bm{C} & \bm{C} & \bm{0} & ...& \bm{0} \vspace{2pt} \\
			\left[^I\bm{p}_{f_2} \times \right]\frac{\partial \bm{\theta}}{\partial \bm{s}} & \bm{0} & \bm{0}& \bm{0} &-\bm{C} & \bm{0} & \bm{C} & ...& \bm{0} \vspace{2pt} \\
			...&...&...&...&...&...&...&...&... \vspace{2pt}\\
			\left[^I\bm{p}_{f_N} \times \right]\frac{\partial \bm{\theta}}{\partial \bm{s}} & \bm{0} & \bm{0}& \bm{0} &-\bm{C} & \bm{0} &\bm{0} &... & \bm{C} \vspace{2pt} \\
			\left[^I\bm{Cv} \times \right]\frac{\partial \bm{\theta}}{\partial \bm{s}} & \bm{0} & \bm{C} & \bm{0} & \bm{0} & \bm{0} &\bm{0}&...&\bm{0} \vspace{2pt} \\
			\bm{0}&\bm{I}&\bm{0}&\bm{0}&\bm{0}&\bm{0}&\bm{0}&...&\bm{0} \vspace{2pt} \\
			\bm{0}&\bm{0}&\bm{0}&\bm{I}&\bm{0}&\bm{0}&\bm{0}&...&\bm{0} \vspace{2pt} \\
			\left[\bm{Cg} \times \right]\frac{\partial \bm{\theta}}{\partial \bm{s}}&\bm{0}&\bm{0}&\bm{0}&\bm{0}&\bm{0}&\bm{0}&...&\bm{0}
		\end{bmatrix}_{(3N+12) \times (3N+15)}\bm{n} =\bm{0}$
	\end{center} 
	
	proof. \begin{center}
		$\begin{matrix}
			d \ ^I\bm{p}_{f} = \bm{K} \\
			d \bm{(Cv)} = \bm{G} \\
			d \bm{b}_g = \bm{J} \\
			d \bm{Cg} = \bm{M} \\
			d \bm{b}_a = \bm{N} \\
			d (\bm{K} \ f_0) =  -d(\bm{Cv}+\left[^I\bm{p}_f \times \right] \bm{b}_g) \\
			d (\bm{K} \ f_1^i) = d(\left[^I\bm{p}_f \times \right] \bm{e}_i) \\
			d (\bm{K} \ f_2^i)=\bm{0} \\
			d (\bm{G} \ f_0) = -d(\left[\bm{Cv} \times \right]\bm{b}_g+\bm{b}_a-\bm{Cg}) \\
			d (\bm{G} \ f_1^i) =d(\left[\bm{Cg} \times \right]\bm{e}_i) \\
			d (\bm{G} \ f_2^i) = \bm{0} \\
			d (\bm{J} \ f_0) = \bm{0} \\
			d (\bm{J} \ f_1^i) = \bm{0} \\
			d (\bm{J} \ f_2^i) = \bm{0} \\
			d (\bm{M} \ f_0) = -d(\left[\bm{Cg} \times \right]\bm{b}_g) \\
			d (\bm{M} \ f_1^i) = -d(\left[\bm{Cg} \times \right]\bm{e}_i) \\
			d (\bm{M} \ f_2^i) = \bm{0} \\
			d (\bm{N} \ f_0) = \bm{0} \\
			d (\bm{N} \ f_1^i) = \bm{0} \\
			d (\bm{N} \ f_2^i) = \bm{0} 
		\end{matrix}$
	\end{center}
	
	Using Theorem 3.1---3.3, the conclusion holds.
	
	$\hfill$ $\Box$
	
	$\textbf{Theorem 3.5.}$ 
	\begin{center}
		$rank(\begin{bmatrix}
			\left[^I\bm{p}_{f_1} \times \right]\frac{\partial \bm{\theta}}{\partial \bm{s}} & \bm{0} & \bm{0}& \bm{0} &-\bm{C} & \bm{C} & \bm{0} & ...& \bm{0} \vspace{2pt} \\
			\left[^I\bm{p}_{f_2} \times \right]\frac{\partial \bm{\theta}}{\partial \bm{s}} & \bm{0} & \bm{0}& \bm{0} &-\bm{C} & \bm{0} & \bm{C} & ...& \bm{0} \vspace{2pt} \\
			...&...&...&...&...&...&...&...&... \vspace{2pt}\\
			\left[^I\bm{p}_{f_N} \times \right]\frac{\partial \bm{\theta}}{\partial \bm{s}} & \bm{0} & \bm{0}& \bm{0} &-\bm{C} & \bm{0} &\bm{0} &... & \bm{C} \vspace{2pt} \\
			\left[^I\bm{Cv} \times \right]\frac{\partial \bm{\theta}}{\partial \bm{s}} & \bm{0} & \bm{C} & \bm{0} & \bm{0} & \bm{0} &\bm{0}&...&\bm{0} \vspace{2pt} \\
			\bm{0}&\bm{I}&\bm{0}&\bm{0}&\bm{0}&\bm{0}&\bm{0}&...&\bm{0} \vspace{2pt} \\
			\bm{0}&\bm{0}&\bm{0}&\bm{I}&\bm{0}&\bm{0}&\bm{0}&...&\bm{0} \vspace{2pt} \\
			\left[\bm{Cg} \times \right]\frac{\partial \bm{\theta}}{\partial \bm{s}}&\bm{0}&\bm{0}&\bm{0}&\bm{0}&\bm{0}&\bm{0}&...&\bm{0}
		\end{bmatrix}_{(3N+12) \times (3N+15)}) = 3N+11$
	\end{center}
	
	i.e. 
	\begin{equation}
		n \in null(O) \iff n\in span\left\{  \begin{bmatrix}
			\bm{0}	& \frac{\partial \bm{s}}{\partial \bm{\theta}}\bm{Cg} \\
			\bm{0}	& \bm{0} \\
			\bm{0}	& -\left[\bm{v} \times \right]\bm{g} \\
			\bm{0}	&  \bm{0} \\
			\bm{I}	&  \left[\bm{g} \times \right] \ ^G\bm{p}_I \\
			\bm{I}	&	\left[\bm{g} \times \right] \ ^G\bm{p}_{f_1} \\
			... & ... \\
			\bm{I} & \left[\bm{g} \times \right] \ ^G\bm{p}_{f_N}
		\end{bmatrix}  \right\} 
	\end{equation}
	
	proof. \begin{center}
		$\left[ \bm{L}_1,\bm{L}_2 .., \bm{L}_N, \bm{L}_{N+1}, \bm{L}_{N+2},\bm{L}_{N+3},\bm{L}_{N+4} \right] \begin{bmatrix}
			\left[^I\bm{p}_{f_1} \times \right]\frac{\partial \bm{\theta}}{\partial \bm{s}} & \bm{0} & \bm{0}& \bm{0} &-\bm{C} & \bm{C} & \bm{0} & ...& \bm{0} \vspace{2pt} \\
			\left[^I\bm{p}_{f_2} \times \right]\frac{\partial \bm{\theta}}{\partial \bm{s}} & \bm{0} & \bm{0}& \bm{0} &-\bm{C} & \bm{0} & \bm{C} & ...& \bm{0} \vspace{2pt} \\
			...&...&...&...&...&...&...&...&... \vspace{2pt}\\
			\left[^I\bm{p}_{f_N} \times \right]\frac{\partial \bm{\theta}}{\partial \bm{s}} & \bm{0} & \bm{0}& \bm{0} &-\bm{C} & \bm{0} &\bm{0} &... & \bm{C} \vspace{2pt} \\
			\left[^I\bm{Cv} \times \right]\frac{\partial \bm{\theta}}{\partial \bm{s}} & \bm{0} & \bm{C} & \bm{0} & \bm{0} & \bm{0} &\bm{0}&...&\bm{0} \vspace{2pt} \\
			\bm{0}&\bm{I}&\bm{0}&\bm{0}&\bm{0}&\bm{0}&\bm{0}&...&\bm{0} \vspace{2pt} \\
			\bm{0}&\bm{0}&\bm{0}&\bm{I}&\bm{0}&\bm{0}&\bm{0}&...&\bm{0} \vspace{2pt} \\
			\left[\bm{Cg} \times \right]\frac{\partial \bm{\theta}}{\partial \bm{s}}&\bm{0}&\bm{0}&\bm{0}&\bm{0}&\bm{0}&\bm{0}&...&\bm{0}
		\end{bmatrix}= \bm{0}$
	\end{center}
	
	$\because$ \begin{center}
		$\left\{\begin{matrix}
			\bm{L}_{N+2}=\bm{0} \\
			\bm{L}_{N+3}=\bm{0} \\
			\bm{L}_{k}\bm{C}=\bm{0} \ (\forall k \le N)\\
			\bm{L}_{N+1}\bm{C}=\bm{0}
		\end{matrix}\right.$ 
	\end{center}
	$\therefore \bm{L}_1=\bm{L}_2=...=\bm{L}_{N+3}=\bm{0}$
	
	$\because \bm{L}_{N+4}\left[\bm{Cg} \times \right]\frac{\partial \bm{\theta}}{\partial \bm{s}} =\bm{0}$, and $\bm{\Psi}$ is diffeomorphism $\therefore$ the left nullspace is $ \gamma \left[\bm{0},\bm{0},\bm{0},\bm{0},\bm{0},...,  \bm{0},(\bm{Cg})^T\right] \  (\gamma \in R)$. Thus, \begin{center}
		$rank(\begin{bmatrix}
			\left[^I\bm{p}_{f_1} \times \right]\frac{\partial \bm{\theta}}{\partial \bm{s}} & \bm{0} & \bm{0}& \bm{0} &-\bm{C} & \bm{C} & \bm{0} & ...& \bm{0} \vspace{2pt} \\
			\left[^I\bm{p}_{f_2} \times \right]\frac{\partial \bm{\theta}}{\partial \bm{s}} & \bm{0} & \bm{0}& \bm{0} &-\bm{C} & \bm{0} & \bm{C} & ...& \bm{0} \vspace{2pt} \\
			...&...&...&...&...&...&...&...&... \vspace{2pt}\\
			\left[^I\bm{p}_{f_N} \times \right]\frac{\partial \bm{\theta}}{\partial \bm{s}} & \bm{0} & \bm{0}& \bm{0} &-\bm{C} & \bm{0} &\bm{0} &... & \bm{C} \vspace{2pt} \\
			\left[^I\bm{Cv} \times \right]\frac{\partial \bm{\theta}}{\partial \bm{s}} & \bm{0} & \bm{C} & \bm{0} & \bm{0} & \bm{0} &\bm{0}&...&\bm{0} \vspace{2pt} \\
			\bm{0}&\bm{I}&\bm{0}&\bm{0}&\bm{0}&\bm{0}&\bm{0}&...&\bm{0} \vspace{2pt} \\
			\bm{0}&\bm{0}&\bm{0}&\bm{I}&\bm{0}&\bm{0}&\bm{0}&...&\bm{0} \vspace{2pt} \\
			\left[\bm{Cg} \times \right]\frac{\partial \bm{\theta}}{\partial \bm{s}}&\bm{0}&\bm{0}&\bm{0}&\bm{0}&\bm{0}&\bm{0}&...&\bm{0}
		\end{bmatrix}_{(3N+12) \times (3N+15)}) = 3N+11$
	\end{center}
	
	$\because  \begin{bmatrix}
		\left[^I\bm{p}_{f_1} \times \right]\frac{\partial \bm{\theta}}{\partial \bm{s}} & \bm{0} & \bm{0}& \bm{0} &-\bm{C} & \bm{C} & \bm{0} & ...& \bm{0} \vspace{2pt} \\
		\left[^I\bm{p}_{f_2} \times \right]\frac{\partial \bm{\theta}}{\partial \bm{s}} & \bm{0} & \bm{0}& \bm{0} &-\bm{C} & \bm{0} & \bm{C} & ...& \bm{0} \vspace{2pt} \\
		...&...&...&...&...&...&...&...&... \vspace{2pt}\\
		\left[^I\bm{p}_{f_N} \times \right]\frac{\partial \bm{\theta}}{\partial \bm{s}} & \bm{0} & \bm{0}& \bm{0} &-\bm{C} & \bm{0} &\bm{0} &... & \bm{C} \vspace{2pt} \\
		\left[^I\bm{Cv} \times \right]\frac{\partial \bm{\theta}}{\partial \bm{s}} & \bm{0} & \bm{C} & \bm{0} & \bm{0} & \bm{0} &\bm{0}&...&\bm{0} \vspace{2pt} \\
		\bm{0}&\bm{I}&\bm{0}&\bm{0}&\bm{0}&\bm{0}&\bm{0}&...&\bm{0} \vspace{2pt} \\
		\bm{0}&\bm{0}&\bm{0}&\bm{I}&\bm{0}&\bm{0}&\bm{0}&...&\bm{0} \vspace{2pt} \\
		\left[\bm{Cg} \times \right]\frac{\partial \bm{\theta}}{\partial \bm{s}}&\bm{0}&\bm{0}&\bm{0}&\bm{0}&\bm{0}&\bm{0}&...&\bm{0}
	\end{bmatrix} \begin{bmatrix}
		\bm{0}	& \frac{\partial \bm{s}}{\partial \bm{\theta}}\bm{Cg} \\
		\bm{0}	& \bm{0} \\
		\bm{0}	& -\left[\bm{v} \times \right]\bm{g} \\
		\bm{0}	&  \bm{0} \\
		\bm{I}	&  \left[\bm{g} \times \right] \ ^G\bm{p}_I \\
		\bm{I}	&	\left[\bm{g} \times \right] \ ^G\bm{p}_{f_1} \\
		... & ... \\
		\bm{I} & \left[\bm{g} \times \right] \ ^G\bm{p}_{f_N}
	\end{bmatrix} = \bm{0}$ 
	\vspace{2pt}
	
	and 
		$rank(\begin{bmatrix}
		\bm{0}	& \frac{\partial \bm{s}}{\partial \bm{\theta}}\bm{Cg} \\
		\bm{0}	& \bm{0} \\
		\bm{0}	& -\left[\bm{v} \times \right]\bm{g} \\
		\bm{0}	&  \bm{0} \\
		\bm{I}	&  \left[\bm{g} \times \right] \ ^G\bm{p}_I \\
		\bm{I}	&	\left[\bm{g} \times \right] \ ^G\bm{p}_{f_1} \\
		... & ... \\
		\bm{I} & \left[\bm{g} \times \right] \ ^G\bm{p}_{f_N}
	\end{bmatrix}) = 4$. The conclusion holds.
	
	$\hfill$ $\Box$
	
	\section{OBSERVABILITY OF LINS}
	
	\ \ \ \ Things become much easier in LINS.
	\begin{equation}
		d\bm{h} = \bm{K}
	\end{equation}
	
	$\textbf{Theorem 4.1.}$ $\bm{Kn}=\bm{0}$
	
	\begin{equation}
		dL_{f_0} \bm{h} = -\left[ ^I\bm{p}_{f} \times \right] \bm{J} -\bm{G} +\left[ \bm{b}_g \times \right] \bm{K}
	\end{equation}
	
	\begin{equation}
		dL_{f_1^i}L_{f_0} \bm{h} = -\left[ \bm{b}_g \times \right] \left[ \bm{e}_i \times \right] \bm{K} - \left[ (\left[ ^I\bm{p}_f \times \right]\bm{e}_i) \times \right] \bm{J} +  \left[ \bm{e}_i \times \right] \bm{G}
	\end{equation}
	
	$\textbf{Theorem 4.2.}$ $\bm{Gn} = \bm{Jn} = \bm{0}$
	
	proof. $\because dL_{f_0} \bm{h} \  \bm{n} = \bm{0}$ \  $\therefore $

	\begin{center}
		$\left[ ^I\bm{p}_{f} \times \right] \bm{J}\bm{n} +\bm{G}\bm{n} = \bm{0}$
	\end{center}
	
	$\because dL_{f_1^i}L_{f_0} \bm{h} \ \bm{n} = \bm{0}$ \ $\therefore$
	
	\begin{center}
		$-\left[ (\left[ ^I\bm{p}_f \times \right]\bm{e}_i) \times \right] \bm{Jn} +  \left[ \bm{e}_i \times \right] \bm{Gn} \ (\forall i)$
	\end{center} i.e.
	\begin{center}
		$- \left[ (\left[ ^I\bm{p}_f \times \right]\bm{e}_i) \times \right] \bm{J}\bm{n} - \left[ (\left[ \bm{J}\bm{n} \times \right] \ ^I\bm{p}_f) \times \right] \bm{e}_i = \bm{0}$ ($\forall i$)
	\end{center}
	 
	 $\because \left[  (\left[\bm{x} \times \right]\bm{y}) \times  \right]\bm{z}+\left[  (\left[\bm{y} \times \right]\bm{z}) \times  \right]\bm{x}+\left[  (\left[\bm{z} \times \right]\bm{x}) \times  \right]\bm{y}=\bm{0} \ (\forall \bm{x},\bm{y},\bm{z}\in R^3)$
	\vspace{2pt}
	
	$\therefore$ $\left[ ^I\bm{p}_f \times \right]\left[ \bm{e}_i \times \right]\bm{Jn}=-\left[ (\left[ \bm{e}_i \times \right]  \bm{Jn}) \times \right] \ ^I\bm{p}_f = \bm{0}$
	
	Because $^I\bm{p}_f \ne 0$, without loss of generality, assume that $p_z \ne 0$. Let $\bm{Jn}=\begin{pmatrix}
		j_1 \\ j_2 \\ j_3
	\end{pmatrix}$.
	
	When $i=3$, $\left[ ^I\bm{p}_f \times \right] \begin{pmatrix}
		-j_2 \\ j_1 \\ 0
	\end{pmatrix} = \bm{0}$. $\therefore j_1 =  j_2 = 0$. When $i=1$, $\left[ ^I\bm{p}_f \times \right] \begin{pmatrix}
		0 \\ -j_3 \\ 0
	\end{pmatrix} = \bm{0}$. $\therefore j_3=0$. $\therefore \bm{Jn}=\bm{Gn}=\bm{0}$
	
	$\hfill$ $\Box$
	
	\begin{equation}
		\begin{matrix}
			dL_{f_0}L_{f_0} \bm{h} = \bm{N} - \bm{M}  \\ 
			-2\left[ \bm{b}_g \times \right] \bm{G} +(\left[ (\left[ ^I\bm{p}_f \times \right]\bm{b}_g) \times \right] - \left[ \bm{b}_g \times \right]  \left[ ^I\bm{p}_f \times \right] + 2\left[ \bm{Cv} \times \right])\bm{J}+\left[ \bm{b}_g \times \right]\left[ \bm{b}_g \times \right]\bm{K}
		\end{matrix}
	\end{equation}
	
	\begin{equation}
		\begin{matrix}
			dL_{f_1^i}L_{f_0}L_{f_0} \bm{h} = \left[ \bm{e}_i \times \right] \bm{M} + 2\left[ \bm{b}_g \times \right]\left[ \bm{e}_i \times \right] \bm{G}-\left[ \bm{b}_g \times \right]\left[ \bm{b}_g \times \right]\bm{K}  \\
			+(2\left[ (\left[ \bm{Cv} \times \right]\bm{e}_i) \times \right]-\left[ \bm{b}_g \times \right]\left[ (\left[ ^I\bm{p}_f \times \right]\bm{e}_i) \times \right]-\left[ (\left[ \bm{b}_g \times \right]\left[ ^I\bm{p}_f \times \right]\bm{e}_i) \times \right])\bm{J}			
		\end{matrix}
	\end{equation}
	
	$\textbf{Theorem 4.3.}$ $\bm{Mn} = \bm{Nn} = \bm{0}$
	
	proof. $\because dL_{f_0}L_{f_0} \bm{h} \ \bm{n} = \bm{0}$ \ $\therefore$
	\begin{center}
		$\bm{Nn}=\bm{Mn}$
	\end{center}
	
	$\because dL_{f_1^i}L_{f_0}L_{f_0} \bm{h} \ \bm{n} = \bm{0}$
	$\therefore$
	\begin{center}
		$\left[\bm{e}_i \times \right]\bm{Mn}=\bm{0} (\forall i)$
	\end{center}
	
	$\therefore\bm{Mn} = \bm{Nn}=\bm{0}$
	
	$\hfill$ $\Box$
	
	$\textbf{Theorem 4.4.}$ $\bm{O \ n} = \bm{0} \iff n\in span\left\{  \begin{bmatrix}
		\bm{0}	& \frac{\partial \bm{s}}{\partial \bm{\theta}}\bm{Cg} \\
		\bm{0}	& \bm{0} \\
		\bm{0}	& -\left[\bm{v} \times \right]\bm{g} \\
		\bm{0}	&  \bm{0} \\
		\bm{I}	&  \left[\bm{g} \times \right] \ ^G\bm{p}_I \\
		\bm{I}	&	\left[\bm{g} \times \right] \ ^G\bm{p}_{f}
	\end{bmatrix}  \right\}$ 
	\vspace{2pt}
	
	proof. Through Theorem 4.1---4.3 and Theorem 3.4---3.5, the conclusion holds.
	
	$\hfill$ $\Box$
	
	\section{PROOF OF OC-VINS}
	
	$\textbf{Lemma 5.1.}$ Denote 
	\begin{equation}
		 \bm{n}_4=\begin{bmatrix}
			 \frac{\partial \bm{s}}{\partial \bm{\theta}}\bm{Cg} \\
			 \bm{0} \\
			 -\left[\bm{v} \times \right]\bm{g} \\
			  \bm{0} \\
			  \left[\bm{g} \times \right] \ ^G\bm{p}_I \\
				\left[\bm{g} \times \right] \ ^G\bm{p}_{f_1} \\
			 ... \\
			\left[\bm{g} \times \right] \ ^G\bm{p}_{f_N}
		\end{bmatrix}
	\end{equation} It holds that
	
	\begin{center}
		$\left[f_0,\bm{n}_4 \right]=\left[f_1^i,\bm{n}_4 \right]=\left[f_2^i,\bm{n}_4 \right]=\bm{0} \ (\forall i)$
	\end{center}
	
	proof. \begin{center}
		$\left[f_1^i,\bm{n}_4 \right] = \begin{bmatrix}
			\frac{\partial \frac{\partial \bm{s}}{\partial \bm{\theta}}\bm{Cg}}{\partial s} \frac{\partial \bm{s}}{\partial \bm{\theta}}\bm{e}_i - \frac{\partial \frac{\partial \bm{s}}{\partial \bm{\theta}}\bm{e}_i}{\partial s} \frac{\partial \bm{s}}{\partial \bm{\theta}}\bm{Cg} \\ \bm{0} \\... \\ \bm{0}
		\end{bmatrix}$
	\end{center} where
	\begin{center}
		$\begin{matrix}
			\frac{\partial \frac{\partial \bm{s}}{\partial \bm{\theta}}\bm{Cg}}{\partial s} \frac{\partial \bm{s}}{\partial \bm{\theta}}\bm{e}_i - \frac{\partial \frac{\partial \bm{s}}{\partial \bm{\theta}}\bm{e}_i}{\partial s} \frac{\partial \bm{s}}{\partial \bm{\theta}}\bm{Cg} \vspace{2pt} \\
			=\left[ \frac{\partial \frac{\partial \bm{s}}{\partial \bm{\theta}}\bm{e}_k}{\partial s}(\bm{Cg})^k +\frac{\partial \bm{s}}{\partial \bm{\theta}} \left[\bm{Cg} \times\right]\frac{\partial \bm{\theta}}{\partial \bm{s}} \right] \frac{\partial \bm{s}}{\partial \bm{\theta}}\bm{e}_i - \frac{\partial \frac{\partial \bm{s}}{\partial \bm{\theta}}\bm{e}_i}{\partial s} \frac{\partial \bm{s}}{\partial \bm{\theta}}\bm{Cg} \vspace{2pt}\\
			=\frac{1}{4}(-\left[\bm{Cg} \times \right]+\bm{s}(\bm{Cg})^T+\bm{s}^T\bm{Cg}\bm{I})(\bm{e}_i-\left[\bm{e}_i \times\right]\bm{s} + \bm{s}^T\bm{e}_i\bm{s})+\frac{1}{2}(\bm{I}+\left[\bm{s} \times \right]+\bm{s}\bm{s}^T)\left[\bm{Cg} \times \right]\bm{e}_i \\
			-\frac{1}{4}(-\left[\bm{e}_i \times\right]+\bm{s}\bm{e}_i^T + \bm{s}^T\bm{e}_i \bm{I})(\bm{Cg}+\left[\bm{s} \times\right]\bm{Cg}+\bm{s}^T\bm{Cg}\bm{s}) \\
			=\frac{1}{4}\left[ -\left[\bm{Cg} \times\right]\bm{e}_i+(\bm{Cg})^T\bm{e}_i\bm{s}-(\bm{Cg})^T\left[\bm{e}_i \times\right]\bm{s} \bm{s}-(\bm{Cg})^T\bm{s} \left[\bm{e}_i \times \right]\bm{s} -\bm{s}^T\bm{e}_i \left[\bm{Cg} \times\right]\bm{s} +2(\bm{s}^T\bm{e}_i)(\bm{s}^T\bm{Cg})\bm{s} \right] \\
			-\frac{1}{4}\left[  -\left[\bm{e}_i \times\right]\bm{Cg}+\bm{e}_i^T\bm{Cg}\bm{s}+\bm{e}_i^T\left[\bm{s} \times\right](\bm{Cg}) \bm{s}-(\bm{Cg})^T\bm{s} \left[\bm{e}_i \times \right]\bm{s} +\bm{s}^T\bm{e}_i \left[\bm{s} \times\right]\bm{Cg} +2(\bm{s}^T\bm{e}_i)(\bm{s}^T\bm{Cg})\bm{s}  \right] \\
			+\frac{1}{2}\left[ \left[\bm{Cg} \times \right]\bm{e}_i +\bm{s}^T\left[\bm{Cg} \times \right]\bm{e}_i \bm{s} +\left[\bm{s} \times \right]\left[\bm{Cg} \times \right]\bm{e}_i \right]\\
			+\frac{1}{4}\left[\bm{s}^T \bm{Cg} \bm{e}_i +\left[\bm{Cg} \times \right]\left[\bm{e}_i \times \right]\bm{s} \right]-\frac{1}{4}\left[ \bm{s}^T \bm{e}_i \bm{Cg} - \left[\bm{e}_i \times \right]\left[\bm{s} \times \right]\bm{Cg} \right]
		\end{matrix}$
	\end{center}
	
	$\because -(\bm{Cg})^T\left[\bm{e}_i \times\right]\bm{s} \bm{s} =-(\bm{Cg})^T\left[\bm{s} \times\right]\bm{e}_i \bm{s} = -\bm{e}_i^T\left[\bm{s} \times \right](\bm{Cg})\bm{s}$
	
	and $\bm{s}^T\left[\bm{Cg} \times \right]\bm{e}_i \bm{s} = -(\bm{Cg})^T \left[\bm{s} \times \right]\bm{e}_i \bm{s}= \bm{e}_i^T\left[\bm{s} \times \right](\bm{Cg})\bm{s}$
	
	$\therefore $\begin{center}
		$\begin{matrix}
				\frac{\partial \frac{\partial \bm{s}}{\partial \bm{\theta}}\bm{Cg}}{\partial s} \frac{\partial \bm{s}}{\partial \bm{\theta}}\bm{e}_i - \frac{\partial \frac{\partial \bm{s}}{\partial \bm{\theta}}\bm{e}_i}{\partial s} \frac{\partial \bm{s}}{\partial \bm{\theta}}\bm{Cg} \vspace{2pt}\\
				=\frac{1}{2}\left[\bm{s} \times \right]\left[\bm{Cg} \times \right]\bm{e}_i+\frac{1}{4}\left[\bm{s}^T \bm{Cg} \bm{e}_i +\left[\bm{Cg} \times \right]\left[\bm{e}_i \times \right]\bm{s} \right]-\frac{1}{4}\left[ \bm{s}^T \bm{e}_i \bm{Cg} - \left[\bm{e}_i \times \right]\left[\bm{s} \times \right]\bm{Cg} \right]
		\end{matrix}$
	\end{center}
	
	$\because \left[\bm{Cg} \times \right]\left[\bm{e}_i \times \right]\bm{s}+\left[\bm{e}_i \times \right]\left[\bm{s} \times \right]\bm{Cg}=-\left[\bm{s} \times \right]\left[\bm{Cg} \times \right]\bm{e}_i$
	
	$\therefore$ \begin{center}
		$\begin{matrix}
			\frac{\partial \frac{\partial \bm{s}}{\partial \bm{\theta}}\bm{Cg}}{\partial s} \frac{\partial \bm{s}}{\partial \bm{\theta}}\bm{e}_i - \frac{\partial \frac{\partial \bm{s}}{\partial \bm{\theta}}\bm{e}_i}{\partial s} \frac{\partial \bm{s}}{\partial \bm{\theta}}\bm{Cg} \vspace{2pt} \\
			=\frac{1}{4}(\left[\bm{s} \times \right]\left[\bm{Cg} \times \right]\bm{e}_i+\bm{s}^T \bm{Cg} \bm{e}_i-\bm{s}^T \bm{e}_i \bm{Cg})=0
		\end{matrix}$
	\end{center}
	
	$\therefore \left[f_1^i,\bm{n}_4 \right]=\bm{0} (\forall i)$
	
	$\because \forall \bm{v} \in R^3, \ \frac{\partial }{\partial s} (\bm{CC}^T\bm{v})=0$, $\therefore \bm{C} \frac{\partial }{\partial s}(\bm{C}^T\bm{v})=-\left[\bm{v} \times \right]\frac{\partial \bm{\theta}}{\partial \bm{s}}$ 
	
	$\therefore \frac{\partial }{\partial s}(\bm{C}^T\bm{v})= -\bm{C}^T\left[\bm{v} \times \right]\frac{\partial \bm{\theta}}{\partial \bm{s}}$
	
	\begin{center}
		$\left[f_0 , \bm{n}_4\right]=\begin{bmatrix}
			\bm{b}_g^k(\frac{\partial \frac{\partial \bm{s}}{\partial \bm{\theta}}\bm{Cg}}{\partial s} \frac{\partial \bm{s}}{\partial \bm{\theta}}\bm{e}_k - \frac{\partial \frac{\partial \bm{s}}{\partial \bm{\theta}}\bm{e}_k}{\partial s} \frac{\partial \bm{s}}{\partial \bm{\theta}}\bm{Cg}) \\ \bm{0} \\-\left[\bm{g} \times \right]\bm{C}^T\bm{b}_a-\bm{C}^T\left[\bm{b}_a \times \right]\bm{Cg} \\ \bm{0}\\\left[\bm{g} \times \right]\bm{v}-\left[\bm{v} \times\right]\bm{g}\\ \bm{0} \\ ...\\ \bm{0}
		\end{bmatrix}$
	\end{center}
	
	$\therefore \left[f_0,\bm{n}_4 \right]=\bm{0}$
	
	\begin{center}
		$\left[f_2^i,\bm{n}_4\right]=\begin{bmatrix}
			\bm{0} \\ \bm{0} \\ \left[\bm{g} \times \right]\bm{C}^T-\bm{C}^T\left[\bm{e}_i \times\right]\bm{Cg} \\ \bm{0} \\ ...\\ \bm{0}
		\end{bmatrix}$
	\end{center} $\therefore \left[f_2^i,\bm{n}_4\right]=\bm{0} \ (\forall i)$
	
	$\hfill$ $\Box$
	
	$\textbf{Theorem 5.1.}$ Under the piecewise-constant input $\bm{\omega}_i$ and $\bm{a}_j$, denote $\bm{f} = f_0 + f_1^i \bm{\omega}_i + f_2^j \bm{a}_j$. The flow of $\bm{f}$ is $\bm{\Phi}_t^{\bm{f}}$. Then $\forall t$ and $\forall \bm{x}_0$, where $\bm{x}_0$ is a state vector, $(\bm{\Phi}_t^{\bm{f}})_* \bm{n}_4(\bm{x}_0)= \bm{n}_4 (\bm{\Phi}_t^{\bm{f}}(\bm{x}_0))$. 
	
	proof. $\frac{d}{d t}(\bm{\Phi}_t^{\bm{f}})_* \bm{n}_4(\bm{\Phi}_{-t}^{\bm{f}}(\bm{x}_0))=\left[\bm{n}_4 , \bm{f}\right](\bm{x}_0) = \left[\bm{n}_4,f_0\right]+\bm{\omega}_i\left[\bm{n}_4,f_1^i\right]+\bm{a}_j\left[\bm{n}_4,f_2^j\right]=\bm{0}$
	
	$\therefore (\bm{\Phi}_t^{\bm{f}})_* \bm{n}_4(\bm{\Phi}_{-t}^{\bm{f}}(\bm{x}_0)) = \bm{n}_4(\bm{x}_0)$, i.e.  $(\bm{\Phi}_t^{\bm{f}})_* \bm{n}_4(\bm{x}_0)= \bm{n}_4 (\bm{\Phi}_t^{\bm{f}}(\bm{x}_0)) \ (\forall \bm{x}_0)$.
	
	$\hfill$ $\Box$
	
	$\textbf{Corollary 5.1.}$ The diffeomorphism is $\bm{\Psi}(\bm{\bar{x}})=\left[\bm{s}(\bm{\theta}), \bm{b}_g, \bm{v}, \bm{b}_a, ^G\bm{p}_I | ^G\bm{p}_{f_1}, ... ,^G\bm{p}_{f_N}\right]$. Denote $\bm{\bar{f}}= (\bm{\Psi}^{-1})_*\bm{f}$, then $(\bm{\Phi}_t^{\bm{\bar{f}}})_*  (\bm{\Psi}^{-1})_*\bm{n}_4(\bm{x}_0)= ((\bm{\Psi}^{-1})_*\bm{n}_4) (\bm{\Phi}_t^{\bm{\bar{f}}}(\bm{\bar{x}}_0)) \ (\forall \bm{\bar{x}}_0 = \bm{\Psi}^{-1}(\bm{x}_0))$, i.e.
	\begin{equation}
		(\bm{\Phi}_t^{\bm{\bar{f}}})_* \begin{bmatrix}
			 \bm{Cg} \\
			\bm{0} \\
			-\left[\bm{v} \times \right]\bm{g} \\
			\bm{0} \\
			\left[\bm{g} \times \right] \ ^G\bm{p}_I \\
			\left[\bm{g} \times \right] \ ^G\bm{p}_{f_1} \\
			... \\
			\left[\bm{g} \times \right] \ ^G\bm{p}_{f_N}
		\end{bmatrix} (\bm{\bar{x}}_0)=\begin{bmatrix}
		\bm{Cg} \\
		\bm{0} \\
		-\left[\bm{v} \times \right]\bm{g} \\
		\bm{0} \\
		\left[\bm{g} \times \right] \ ^G\bm{p}_I \\
		\left[\bm{g} \times \right] \ ^G\bm{p}_{f_1} \\
		... \\
		\left[\bm{g} \times \right] \ ^G\bm{p}_{f_N}
		\end{bmatrix} ( \bm{\Phi}_t^{\bm{\bar{f}}}(\bm{\bar{x}}_0) )		
	\end{equation}	
	
			proof. $\because (\bm{\Psi}^{-1})_* \left[\bm{n}_4, \bm{f}\right] = \left[(\bm{\Psi}^{-1})_*\bm{n}_4, (\bm{\Psi}^{-1})_*\bm{f}\right]=0$, $\therefore$ the conclusion holds.
			
			$\hfill$ $\Box$

\end{document}